\documentclass[preprint,preprintnumbers,amsmath,amssymb]{revtex4}
\pdfoutput=1
\usepackage[pdftex]{graphicx}

\begin{document}

\title{
A constrained, total-variation minimization algorithm for low-intensity
X-ray CT}

\author{Emil Y. Sidky}
\email{sidky@uchicago.edu}

\author{Yuval Duchin}

\author{Xiaochuan Pan}
\email{xpan@uchicago.edu}
\affiliation{%
University of Chicago \\
Department of Radiology \\
5841 S. Maryland Ave., Chicago IL, 60637
}%

\author{Christer Ullberg}
\affiliation{%
XCounter AB\\
Sv\"{a}rdv\"{a}gen 11 \\
SE-182 33 Danderyd, Sweden
}%

\date{\today}

\begin{abstract}
PURPOSE:\\
\noindent
We develop an iterative image-reconstruction algorithm for
application to low-intensity computed tomography (CT) projection data, which
is based on constrained, total-variation (TV) minimization. The algorithm
design focuses on recovering structure on length scales comparable to
a detector-bin width.

METHOD:\\
\noindent
Recovering the resolution on the scale of a detector bin, requires
that pixel size be much smaller than the bin width.  The resulting
image array contains many more pixels than data, and this undersampling
is overcome with a combination of Fourier upsampling of each projection
and the use of constrained, TV-minimization, as suggested by
compressive sensing. The presented pseudo-code for solving
constrained, TV-minimization is designed to yield an accurate
solution to this optimization problem within 100 iterations.

RESULTS:\\
\noindent
The proposed image-reconstruction algorithm is applied to a low-intensity scan of a rabbit
with a thin wire, to test resolution. The proposed algorithm is compared
with filtered back-projection (FBP). 

CONCLUSION:\\
\noindent
The algorithm may have some advantage
over FBP in that the resulting noise-level is lowered at equivalent contrast levels
of the wire.
\end{abstract}


\maketitle

\section{Introduction}

Motivated by the desire to reduce dose \cite{McCollough:09} and the ever-climbing availability
of cheap computation power, much effort has been directed to developing iterative
image reconstruction (IIR) for application in X-ray CT \cite{erdogan1999ordered,qi2006iterative,PanIP:09}.
When considering a fixed total X-ray
dose for a given scan, there is a trade-off between intensity-per-view and number-of-views.
Much of the recent work on IIR based on a constrained, $\ell_1$ or total variation (TV) optimization
problem has explored the sparse-view end of this trade-off
\cite{li2002accurate,SidkyTV:06,chen2008prior,song2007sparseness,sidky2008image,sidky2009enhanced,PanIP:09,SidkyPC:10,jia2010gpu,Choi:10,SidkyPC:10,Bergner:10}.
While the 
low-intensity/many-views end of this spectrum is generally dealt with by employing
filtered back-projection (FBP) with regularization or IIR based on a maximum-likelihood
principle.  In this work, we would like to extend IIR based on constrained, TV-minimization
to handle CT data with many projections and low-intensity (high-noise) per projection.

The use of constrained, TV-minimization derives from recent theory in compressive
sensing (CS) \cite{candes2006robust,candes2006stable,candes2008introduction},
where certain sparsely sampled linear systems can be inverted accurately
when the underlying object has an approximately sparse gradient-magnitude image.
The CS-motivated optimization problem
appears to be effective
for accurate image reconstruction from sparse-view data as evaluated by many image
quality metrics \cite{Bian:10}. The obvious
question, now, is why would we want to extend constrained, TV-minimization to CT data with
many projections?  The answer is that no matter how many projections a CT data set contains, there
may always be an issue with view-angle under-sampling. Particularly in diagnostic X-ray CT, the bar for 
image quality is quite high; it is often expected that detail on the scale of a single detector
bin (0.1-0.25 mm) will be visible. At such scales, images of structures are often degraded
due to the fact that standard CT scans -- even with ~1000 projections --
contain too few views.  Further evidence of under-sampling in CT practice is that industry has
developed a hardware solution,
which is a X-ray source with a flying focal-spot to effectively double the number of projections
\cite{flohr2004performance,kachelriess2006flying}.
By developing constrained, TV-minimization for low-intensity/many-view data, we seek a software
solution to this problem. The technical problem to be overcome is how to effectively deal
with noisy projection data in the resulting IIR algorithm.

The main goal of this article is to report a constrained, TV-minimization IIR algorithm for
low-intensity/many-view CT projection data. The algorithm is derived from a framework we have
been developing where constrained, TV-minimization is solved with a combination of steepest-descent (SD),
to reduce image TV, and projection onto convex sets (POCS), to enforce data-error and other image constraints.
As the step-size of the SD component of the algorithm is adaptively adjusted, the algorithm framework is
referred to as adaptive SD-POCS (ASD-POCS). The particular flavor of ASD-POCS presented here is designed
to solve constrained, TV-minimization accurately in a reasonable number of iterations (~100 iterations).
This ASD-POCS algorithm is demonstrated with an XCounter CT scan of a rabbit with a thin wire taped to
the outside of the sample holder. The data are low-intensity and contain 1878 projections with a 2266x64
bin detector at a resolution of 0.1 mm. The thin wire provides a good test for the image reconstruction algorithm.
For our purpose, we take the middle row on the detector from this data set and focus on 2D fan-beam CT reconstruction
with 1878 projections on a 2266-bin linear detector array.
In Sec. \ref{sec:IIRsampling} we discuss in detail a fundamental issue of sampling for IIR; in Sec. \ref{sec:asdpocs}
we present the ASD-POCS algorithm for low-intensity CT; and in Sec. \ref{sec:results} the algorithm is applied
to the rabbit scan.

\section{Sampling and image representation for high-resolution CT imaging}
\label{sec:IIRsampling}

For FBP, which is an analytic-inverse-based image reconstruction algorithm, the data sampling requirements
are guided by the fact one needs a good estimate of the continuous projection data. Whether done explicitly or
not, the discretely sampled X-ray transform is interpolated to a continuous function, then fed into an analytic-inverse
formula for the X-ray transform.  In the theory for CT sampling, there is much discussion about satisfying
a Nyquist sampling condition for the data, but in practice this condition is used only as an
estimate of resolution for a given CT system.  Often objects being scanned in CT have edge-discontinuities, which
violate the band-limited requirement of Nyquist sampling. Furthermore, most implementations of FBP use
linear-interpolation in the filtering and back-projection integrals instead of the sinc-interpolation called for
by the sampling theorem. In any case, the CT sampling issue boils down to how well the interpolated data function
matches the continuous projection of the underlying object function. The
FBP image can be displayed on a grid of any size, because FBP provides a closed-form expression for the image
in terms of the data. The accuracy of this image, however, depends on the accuracy of the interpolation of the data function.

For IIR, which uses a discrete data model, the image resolution depends on two things: (1) the expansion set used
to represent the image, and (2) the number of measurements available to determine the expansion coefficients.
The first step is to design
an expansion set for the underlying object function. For the present work, we choose image pixels as this
expansion set. Fixing the expansion set, the next step to understanding the sampling is to determine if there
are enough ray-integration measurements to specify the expansion coefficients. The required amount of data to
determine a unique image depends on the number of expansion elements.
To explain this sampling issue for IIR more concretely, we use the configuration of the XCounter CT of a rabbit-plus-wire.

The projection of the rabbit is confined to the middle 1266 bins of the detector so the data size is effectively
1878 views by 1266 bins with each bin measuring 0.1 mm in width.  We would like to resolve structure within a
(0.1 mm)$^2$ region, and as a result the pixels representing the image must be much smaller than this 0.1 mm square
\cite{zbijewski2004characterization}.
Say we choose pixels of size 0.025 mm so that the 0.1 mm square has 16 sub-elements.  It turns out the the support
of the rabbit can be covered by a 4096x4096 array of pixels of size (0.025 mm)$^2$. With this choice of
parameters, the number of pixels is much larger than the number of measurements. If instead we had decided to
use (0.1 mm)$^2$ pixels, the discrete data model would not be an under-determined linear system.
It is clear, however, that the data model will always be under-determined
if the pixel-size is chosen to be smaller than the detector bin width. Using alternative
basis functions does not resolve this dilemma; whenever it is desirable to recover structure on the scale
of a detector bin, there will be many more expansion elements than measurements.

Within the framework of optimization-based image reconstruction, such under-sampling problems are resolved
by the exploitation of some kind of prior knowledge.  One possible choice is to exploit sparsity in the
gradient magnitude image, and employ constrained, TV-minimization. Mathematically, the constraints of having
to agree with the data and image non-negativity yields a multiplicity of images. But there will in general
be one image, with in this feasible sub-set, that has a minimum image TV. While constrained, TV-minimization
has proved useful for angular under-sampling, it may not be as effective when {\it both} the scanning angle and detector
bin direction are under-sampled, as is the case here. 

A possible solution to the problem of how to employ a super-resolution grid of pixels comes from analyzing
the sampling for FBP.
CT sampling is not uniform and the limiting
factor is usually the angular sampling rate. As a prior on the system, we can assume that the sampling along the
projection does satisfy the Nyquist sampling condition.  If this is the case, we can generate more samples
by Fourier interpolation, zero-padding the projection's Fourier transform,
to augment the data set to 1878 views by 5064 (4$\times$1266) bins. With this set of data, we are no
longer undersampled on the direction along the detector. Now, we can exploit sparsity in the gradient
magnitude image, by basing the IIR algorithm on constrained, TV-minimization. And we can expect this
strategy to be successful, as constrained, TV-minimization has been demonstrated to be effective against
angular under-sampling. Although, we have chosen a factor of 4, the method can be extended to even larger
sub-sampling factors because, under the assumption of Nyquist sampling along
the detector, the number of samples per projection can scale with the pixel grid size.
Another extension of this idea is to use other methods to interpolated the projections, for example, linear
interpolation.

\section{The ASD-POCS algorithm for low-intensity CT}
\label{sec:asdpocs}

Up until now, we have not addressed the issue of the high noise-level at each projection. In this work,
we do not explicitly incorporate a noise model into the design of the IIR algorithm. Instead the consideration
of noise is more of a practical issue in that it turns out to be difficult to solve the constrained, TV-minimization
problem with a large number of views and a high noise level per view.
The specific data model for our system is a linear equation:
\begin{equation}
\label{DDmodel}
\tilde{g} = X \vec{f},
\end{equation}
where $\tilde{g}$ represents the augmented projection data, in this case a vector of length $1878 \times 5064$;
$\vec{f}$ is a vector of pixel values on the super-resolution grid, here $4096 \times 4096$; and $X$ is the ray-driven
model of the X-ray transform where system matrix element is the intersection length of a given ray through a given pixel.
An IIR algorithm based on constrained, TV-minimization aims at solving:
\begin{equation}
\label{TVmin}
\vec{f}^* = \text{argmin} \| \vec{f} \|_\text{TV} \text{  such that  } |X \vec{f} - \tilde{g}|^2 \le \epsilon^2
\; \; \; \vec{f} \ge 0,
\end{equation}
where $\| \vec{f} \|_\text{TV}$ is the sum over the gradient magnitude image; and $\epsilon$ is a data-error tolerance parameter.
Because there will be no image that exactly reproduces the data, due to noise and other physical
factors, there will be a non-zero minimum data-error tolerance $\epsilon_\text{min}$. This optimization problem
can be difficult to solve for our system; especially, because we are interested in values of $\epsilon$ near
$\epsilon_\text{min}$.  An IIR algorithm, however, can be designed to solve this problem efficiently by converting
Eq. (\ref{TVmin}) to an equivalent least-absolute-shrinkage-and-selection-operator (LASSO) optimization problem \cite{Wright:07}.

In the LASSO form, the term representing the data error goes into the objective function and the image TV is swapped out
as a constraint:
\begin{equation}
\label{LASSO}
\vec{f}^* = \text{argmin} |X \vec{f} - \tilde{g}|^2 \text{  such that  }   \| \vec{f} \|_\text{TV} \le t_0
\; \; \; \vec{f} \ge 0,
\end{equation}
where the parameter $t_0$ is the maximum allowed image TV. This parameter replaces the $\epsilon$ from
Eq. (\ref{TVmin}).  To solve Eq. (\ref{TVmin}), one selects a $t_0$, then solves Eq. (\ref{LASSO}).
The value of $\epsilon$ corresponding to $t_0$ is found be simply evaluating the objective function
for $\vec{f}^*$.  This optimization problem is more amenable for algorithm design for a few reasons:
(1) We are interested in low $\epsilon$ which corresponds to high $t_0$ -- thus the feasible set of images is large;
(2) The initial estimate of a zero image has zero image-TV and is thus in the feasible set from the beginning; and
(3) It is efficient to project images into the feasible set because the constraints can be evaluated quickly
for a given image estimate. The optimality conditions for Eq. (\ref{LASSO}) fall into two cases:
First, if $t_0$ is chosen too large then the image-TV constraint is satisfied with a strict inequality;
the image is non-negative; 
and the gradient of the data-residual objective function, masked by the image estimate support, has zero length.
The masking by the image support comes from the non-negativity constraint \cite{sidky2008image}. Second, the more useful
case, which is equivalent to Eq. (\ref{TVmin}), is when the image-TV constraint is active and is therefore
satisfied with equality. In this case, we define an angle $\alpha$
between the gradient of the data residual, masked by the image support, and the gradient of the image TV,
also masked by the image support. At optimality this angle should be 180$^\circ$ or $\cos \alpha = -1,$ and of course
the image should be non-negative. This condition is derived and described in more detail in Ref. \cite{sidky2008image}.
The condition $\cos \alpha = -1$ is a very sensitive test, and is therefore quite useful for the present
purposes, because we aim at solving Eq. (\ref{LASSO}), accurately.
The use of a data error plot with iteration number, as is often done, does not indicate convergence
because we are solving an under-determined problem and there is a large multiplicity of images for a given
data residual.

The algorithm designed to solve Eq. (\ref{LASSO}) is an ASD-POCS algorithm, in that SD with an adaptive
step-size is used to lower image TV and POCS is employed to lower the data residual objective function.
The pseudo-code is:

\begin{tt}
\begin{tabbing}
~~~~~~~\=~~~\=~~~\=~~~~~\=~~~~~~~~~~~~~~~~~~~~~~~~~~~~~\=~~~~~\=\\
\noindent
1:  \> $\beta := 1.0; \; \beta_\text{red} := 0.7; \; \beta_\text{min} := 10^{-5} $ \\
2:  \> $\rho_\text{min} := 1.1; \; \rho_\text{max} := 2.0; $ \\
3:  \> $\gamma_\text{red}:=0.8$ \\
4:  \> $\vec{f} := 0$ \\
5:  \> while $\beta \ge \beta_\text{min}$ do \\
6:  \>     \> $\vec{f}_0 :=  \vec{f}$ \\
7:  \>     \> for $j = 1, N_d$ do $\vec{f} :=  \vec{f} +\beta
\vec{X}_j \frac{g_j - \vec{X}_j \cdot \vec{f} }{\vec{X}_j \cdot \vec{X}_j}$ \\
8:  \>     \> $\vec{f} := P(\vec{f})$ \\
9:  \>     \> $\vec{p} := \vec{f} - \vec{f}_0  $ \\
10: \>     \> $\rho := S[ TV(\vec{f}_0 + \rho \vec{p})-t_0=0 , \rho ]  $ \\
11: \>     \> $\rho := \min (\rho,\rho_\text{max}) $ \\
12: \>     \> $\vec{f} := \vec{f}_0 + \rho \vec{p} $ \\
13: \>     \> if $TV(\vec{f})=t_0$ and $\rho<\rho_\text{min}$ then \\
    \>     \>     \>  $\beta := \beta*\beta_\text{red}$ \\
14: \>     \> $\vec{f}_\text{res} := \vec{f}$ \\
15: \>     \> $dp := |\vec{f} - \vec{f}_0| $ \\
16: \>     \> if $TV(\vec{f})=t_0$ then \\
17: \>     \>     \> $\vec{df} := \nabla_{\vec{f}}  TV(\vec{f}) $ \\
18: \>     \>     \> $\hat{df} := \vec{df}/|\vec{df}|$ \\
19: \>     \>     \> $\vec{f}^\prime := \vec{f}- dp* \hat{df}$ \\
20: \>     \>     \> $\vec{f}^\prime := P(\vec{f}^\prime)$ \\
21: \>     \>     \> $\gamma := 1.0$ \\
22: \>     \>     \> while  $TV(\vec{f}^\prime)>t_0$ do \\
23: \>     \>     \>       \> $\gamma := \gamma * \gamma_\text{red}$ \\
24: \>     \>     \>       \> $\vec{f}^\prime := \vec{f}- \gamma dp* \hat{df}$ \\
25: \>     \>     \>       \> $\vec{f}^\prime := P(\vec{f}^\prime)$ \\
26: \>     \>     \> end while  \\
27: \>     \>     \> $\vec{f} := \vec{f}^\prime$ \\
28: \>     \> end if  \\
29: \> end while\\
30: \> return $\vec{f}_\text{res}$
\end{tabbing}
\end{tt}

The general idea of the algorithm is to start with a zero image estimate which obviously
satisfies non-negativity and the image TV constraints. A POCS-step is computed which
reduces the data error while maintaining non-negativity. This step is scaled so that
the image estimate goes to the boundary of the feasible space $TV(\vec{f}) = t_0$.
A single SD-step on the image TV is then taken with a line-search to ensure that the 
image TV is reduced, taking the image estimate to the interior of the TV
constraint. The image estimate after a single loop of POCS and SD will on average
have a lower data residual and will remain in the interior of the TV constraint.
Repetition of this loop will slide the image along the boundary of the TV constraint,
maintaining non-negativity, until a minimum data error is reached.

The data error reduction happens at line 7 with the standard algebraic reconstruction technique (ART)
loop, where $\vec{X}_j$ is the row of the system matrix yielding an estimate for the ray integration
corresponding the data element $g_j$ and $N_d$ is the number of ray measurements in the augmented
data set; for the results below $N_d= 1878 \times 5064$.  
Line 8 enforces non-negativity
with the function $P(\cdot)$, which puts zeros in any component of the argument that are negative;
lines 7 and 8 together are POCS. The relaxation factor $\beta$
at line 7 starts at a value of 1.0 and is reduced aggressively by a factor $\beta_\text{red}$
defined at line 1. Termination of the program is based on testing $\beta$ against a minimum
value at line 5.  The program is designed so that the current image estimate $\vec{f}$
before ART at line 7 satisfies the image TV constraint, $TV(\vec{f})<t_0$, with inequality.
The function $S[ TV(\vec{f}_0 + \rho \vec{p})-t_0=0 , \rho ]$ solves the non-linear equation in the first
argument of $S[\cdot,\cdot]$ for $\rho$, the second argument of $S[\cdot,\cdot]$.
If there is no solution, the value of $\rho$ is selected that 
minimizes the difference magnitude on the left-hand side; if this is the case the resulting
image TV will be less than $t_0$ instead of greater. The scale factor $\rho$ needed to 
bring the TV of the image estimate to $t_0$. This factor is bounded above at line 11
by the value 2 in order that the ART-step does not increase the data error. The ART-step
with a scale factor is added to the image estimate at line 12.  There are two conditions
for reducing the relaxation factor at line 13.
The first condition checks if the POCS step with scaling could successfully bring
the image estimate to the boundary of the feasible space. This check is necessary,
because it is possible that the relaxation factor is reduced too fast. If this is the case
the image estimate will remain in the interior of the TV constraint after POCS, and in this
case we do not want to reduce the relaxation factor further.
The second checks if the scale factor $\rho$ is
below a minimum value. This test effectively adjusts the ART-step size quickly to the problem
at hand.
The image estimate
is stored in $\vec{f}_\text{res}$ at line 14; this will be the final image on termination of the
program. The magnitude of the image change due to POCS, $dp$, is computed at line 15 for use in
the adaptive SD on the image TV. The SD portion of the program at lines 16-28 are executed only
if the POCS step successfully reached an image TV of $t_0$. If this is not the case, the
image TV will be less than $t_0$ and we do not want to reduce it further. The adaptive
aspect of the SD-step is seen at line 20 where the step search is started with the value of $dp$.
The choice of algorithm parameters at lines 1-3 are what we used for the following results section.

The critical parameters are $\beta_\text{red}$ and $\rho_\text{min}$. If $\beta_\text{red}$ is
chosen too small then the program terminates too quickly, well before convergence. Likewise,
higher values of $\rho_\text{min}$ cause the relaxation factor to be reduced more often.
A value of $\rho_\text{min}$ should be greater than or equal to 1.0. With higher values reducing
the number of iterations.  Critical is the $\cos \alpha = -1$ test. A good strategy is to
start with aggressive parameters, where it will be clear whether or not convergence can be
achieved within 10-20 iterations. If not, then $\beta_\text{red}$ can be increased or
$\rho_\text{min}$ can be reduced.  $\cos \alpha$ will in general not reach -1.0, but values
below -0.5 generally indicate proximity to the solution. Because of the high dimensionality
of the image coefficient vector $\cos \alpha< -0.5$ indicates a small error per pixel, if the
error is distributed evenly over all pixels.

We stress that this form of the ASD-POCS algorithm is designed for IIR in the situation
where the desired operating range for image regularization is relatively weak and the
data error tolerance is near its minimum possible value. Qualitatively, the resulting
images will still have speckle noise, albeit at a lower level. If images are desired,
which are regularized to the point where the speckle noise is removed, then it is better
to use the basis pursuit, Eq. (\ref{TVmin}), optimization problem to design an algorithm,
because the feasible region for the LASSO problem shrinks while that of the basis pursuit
expands.

Finally, because the goal of the algorithm is an accurate solution to Eq. (\ref{LASSO}), the
resulting images can be regarded as a function of only the scanning parameters and $t_0$.
The details of the algorithm, both particular parameter settings of methods for reducing
data error or image TV, are only important for algorithm efficiency and they do not affect
the final image.
On the other hand, this means we must take the optimality conditions seriously. In the
results section, we give a sense of image dependence on $\cos \alpha$ to demonstrate
that the error in solving the LASSO equation is well below the visual threshold of detecting
a difference in the image. A question arises on how to choose $t_0$.  As the application, here,
is to perform image reconstruction which has a lower noise level than that obtained by standard
FBP, the FBP image itself provides a reference value for $t_0$.  In the results, below, we
show images for different values of $t_0$, and the optimal value will depend on imaging task.

\section{Results: LASSO-form ASD-POCS applied to a rabbit scan with a thin wire}
\label{sec:results}

We use the rabbit-scan with a thin wire to demonstrate the LASSO-form of the ASD-POCS
algorithm on finely sampled projection data with low X-ray intensity. The first set
of results are aimed at illustrating various points about the algorithm itself; we discuss
algorithm convergence and the need to perform up-sampling in the projection data.
The second set of results compares the LASSO-form ASD-POCS algorithm with a standard
FBP algorithm over a range of image regularizations.

\subsection{Illustration of the LASSO-form ASD-POCS IIR algorithm: convergence and upsampling}

\begin{figure}[!t]
\begin{minipage}[b]{0.8\linewidth}
\centering
\centerline{\includegraphics[width=8cm,clip=TRUE]{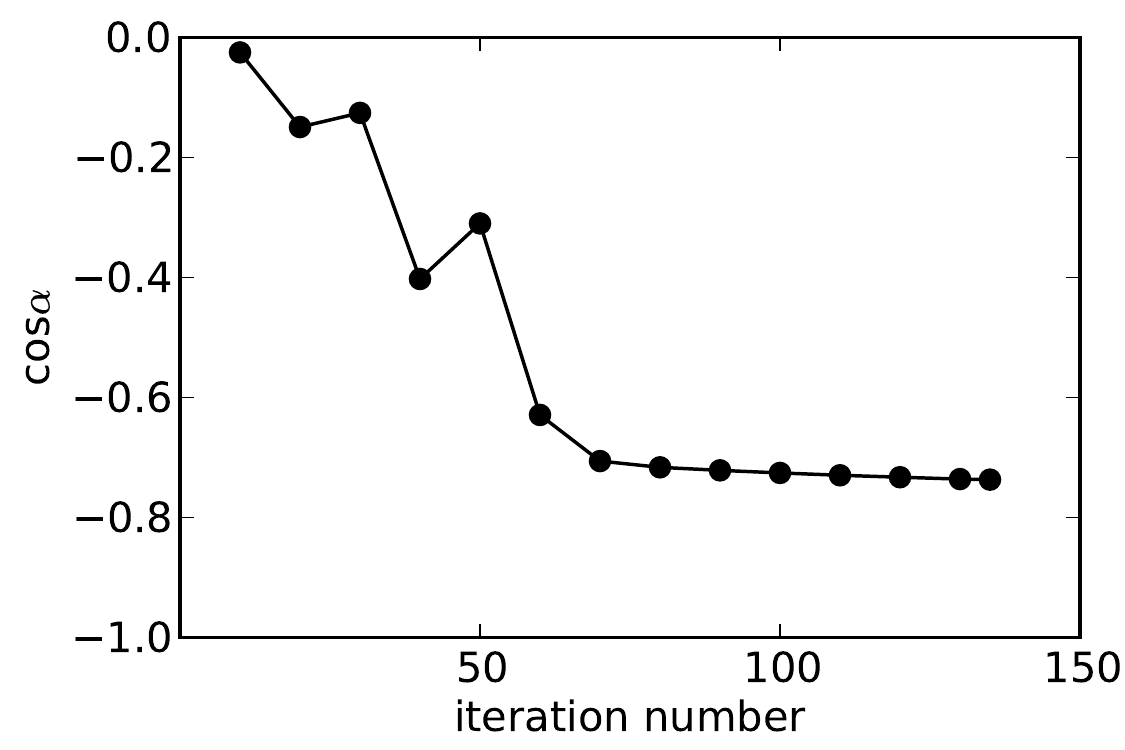}}
\end{minipage}
\caption{
Evolution of $\cos \alpha$ with iteration number for an example
run of the LASSO-form ASD-POCS algorithm.
\label{fig:cosa}}
\end{figure}

As noted above the size of the reconstruction problem solved, here, is relatively large
for a 2D CT system. The image array consists of 4096x4096 pixels and the upsampled data
contain 1878$\times$5064 measurements resulting in a system matrix of size
$ \approx 10^7 \times (1.6 \times 10^7)$.  Fortunately, computations on a commodity graphics
processing unit (GPU), originally introduced to the medical imaging community
by Mueller {\it et al.} \cite{Mueller:05}, make possible a substantial acceleration
by approximately a factor of 10, for our case.
Even though we have implemented the ART-step of line 7 in CUDA using a Tesla C1060
GPU, this step still takes a few minutes of computation time.  Thus efficiency of the
ASD-POCS algorithm itself is still important.

\begin{figure}[!t]
\begin{minipage}[b]{0.8\linewidth}
\centering
\centerline{\includegraphics[width=8cm,clip=TRUE]{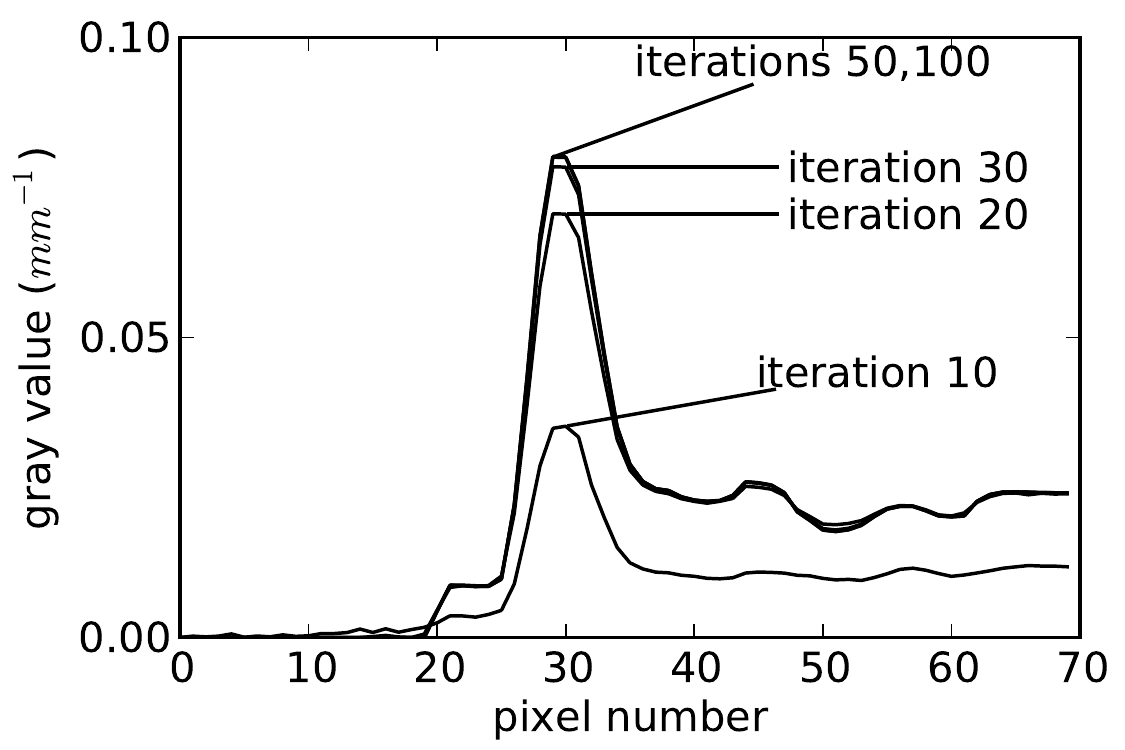}}
\end{minipage}
\caption{
Profile through wire for different iteration numbers of an example run
of the LASSO-form ASD-POCS algorithm.
\label{fig:iterprof}}
\end{figure}

To demonstrate convergence of one of the ASD-POCs reconstructions, we show 
$\cos \alpha$ as a function of iteration number in Fig. \ref{fig:cosa}.
We point out that all other constraints of Eq. (\ref{LASSO}), positivity and the 
TV-bound are satisfied. Within tens of iterations
$\cos \alpha$ drops below -0.5, a value which on the face of it seems rather far from the
truly converged value of -1.0. But the image space, here, is large -- $16 \times 10^6$ pixels.
With such high dimensionality, a value of -0.5 results in a fairly accurate image.
For example, suppose that the error from the true solution is a random image following
an independent uniform Gaussian distribution. One can show that the average deviation per pixel
from the true solution is 0.04\% for $\cos \alpha = -0.5$. Of course, we do not expect that
the error image follows this model, but at least this gives a sense of the meaning of $\cos \alpha$.
As an independent demonstration of convergence, we show a series of one dimensional
profiles corresponding to different iteration numbers through the wire in the contained in the subject
in Fig. \ref{fig:iterprof}.
The difference in the profiles between 50 and 100 iterations is imperceptible, even though $\cos \alpha$
drops from -0.31 to -0.74 over this range. The difference images as a function of iteration seems
to agree with the Gaussian error model. Although we show only one example here, we have verified similar
convergence properties for this version of ASD-POCS for numerous scanning conditions.
Thus, we claim that the images shown in this article are visually indistinguishable from the
true solution of Eq. (\ref{LASSO}) for the gray scale ranges shown.

\begin{figure}[!t]
\begin{minipage}[b]{0.8\linewidth}
\centering
\centerline{\includegraphics[width=8cm,clip=TRUE]{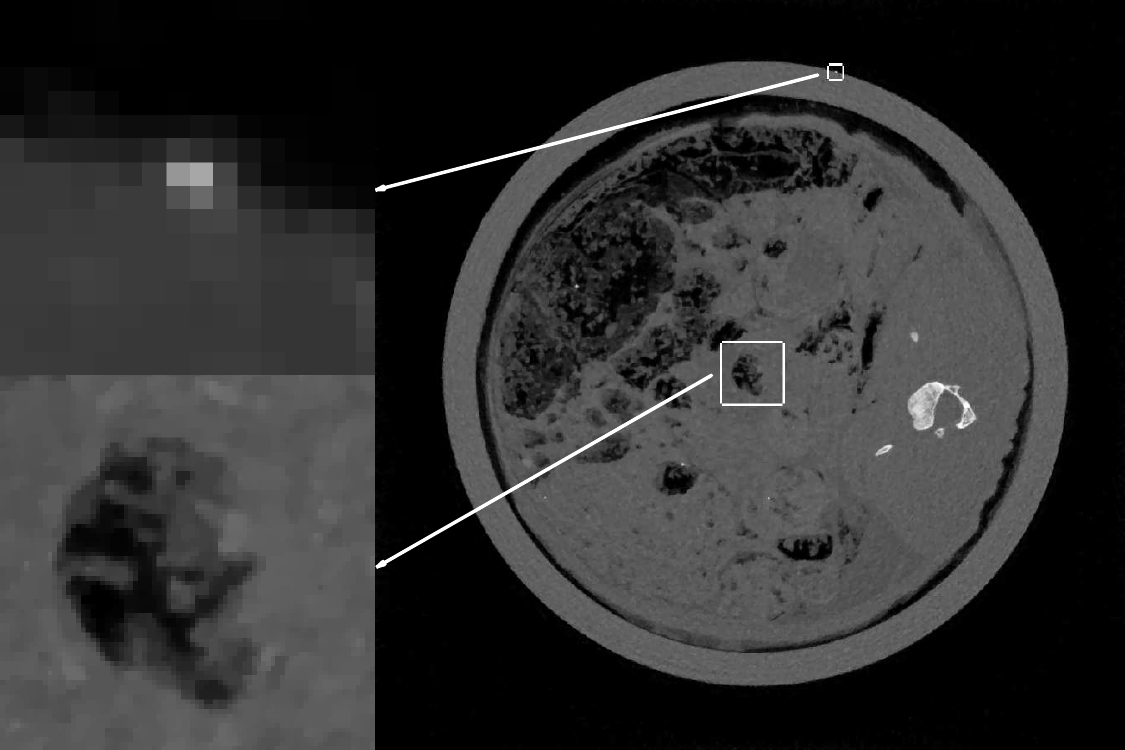}}
\vskip 0.25cm
\centerline{\includegraphics[width=8cm,clip=TRUE]{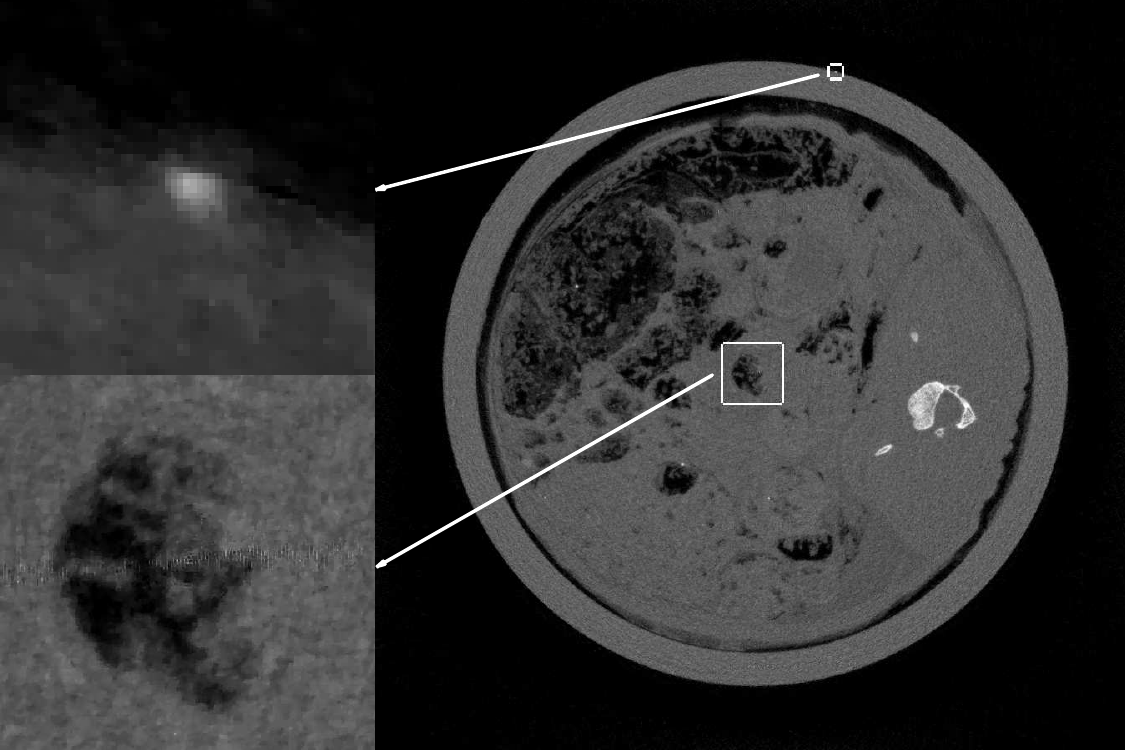}}
\vskip 0.25cm
\centerline{\includegraphics[width=8cm,clip=TRUE]{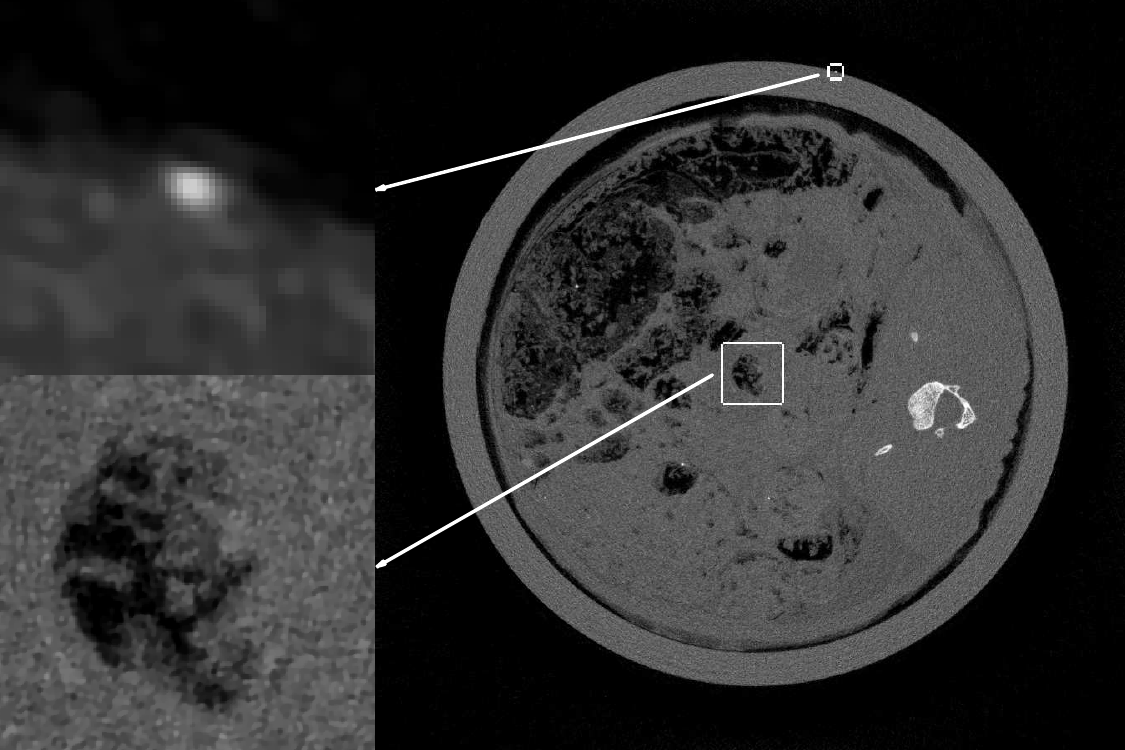}}
\end{minipage}
\caption{Top: ASD-POCS reconstruction from 1878$\times$1266 data set to a 1024$\times$1024
image array. Middle: ASD-POCS reconstruction from same data set to a 4096$\times$4096 image array.
Bottom: ASD-POCS reconstruction from 1878$\times$5064 upsampled data set to a  4096$\times$4096 image array.
For each image the gray scale is [0, 0.06] $mm^{-1}$ accept for the top, left ROI containing the cross-section
of the wire, which is displayed in a window of [0,0.1] $mm^{-1}$.
\label{fig:upsampling}}
\end{figure}

To demonstrate the importance of the projection data upsampling to squeeze out the resolution contained
in the data, we compare images for three cases shown in Fig. \ref{fig:upsampling}.
First, we show the ASD-POCS image obtained when the image array is 1024$\times$1024
at a pixel size is (0.1 mm)$^2$, the same as the
detector bin size, and no upsampling of the data is performed. Second, we increase the image array
to 4096$\times$4096, or
equivalently, decrease the pixel size to (0.025 mm)$^2$, and no upsampling of the data is performed.
Finally, the 4096$\times$4096 image array is employed with each projection being upsampled
by a factor of four. All computations are done
at equivalent $t_0$. The small image array is clearly not up to the task as the wire appears 
as a single square. Moreover, the overall impression of the image appears blotchy -- a criticism that
has been leveled against the use of TV in many articles. Going to the larger array, without data upsampling,
improves the image, but the reconstruction is a difficult inversion problem in this case
because the undersampling factor is not small and both
dimensions of the data space are undersampled relative to the pixel array. Inspection of the
image shows some peculiarities in the noise pattern, where widely-spaced, large-amplitude, salt-and-pepper noise appears,
and artifacts are clearly visible in the lower left panel where gaps between the measurement rays cause some
striping.
High values of the noise pattern could be mistaken as tiny micro-calcifications. Finally, the high resolution
array combined with the upsampled data appears to properly reconstruct the wire while not introducing
a strange noise pattern or artifacts.

We point out here that the strategy of upsampling the data is not the only possibility of improving the
condition-number of the discrete imaging model.  A strip integration model for the projection data, where the
extended X-ray source-spot and detector-bin are taken into account, would likely yield decent image quality with
the large image array. But this seemingly more realistic model does not necessarily model the CT system
better than the present upsampling approach, because an accurate model of the physics would include a non-linear
averaging of the rays in the strip and not just straight summation of the rays contributing to a single measurement
\cite{Zou:04}. We leave the investigation of alternate projection models to future work.

\subsection{Image regularization through varying $t_0$}

\begin{figure}[!t]
\begin{minipage}[b]{0.8\linewidth}
\centering
\centerline{\includegraphics[width=8cm,clip=TRUE]{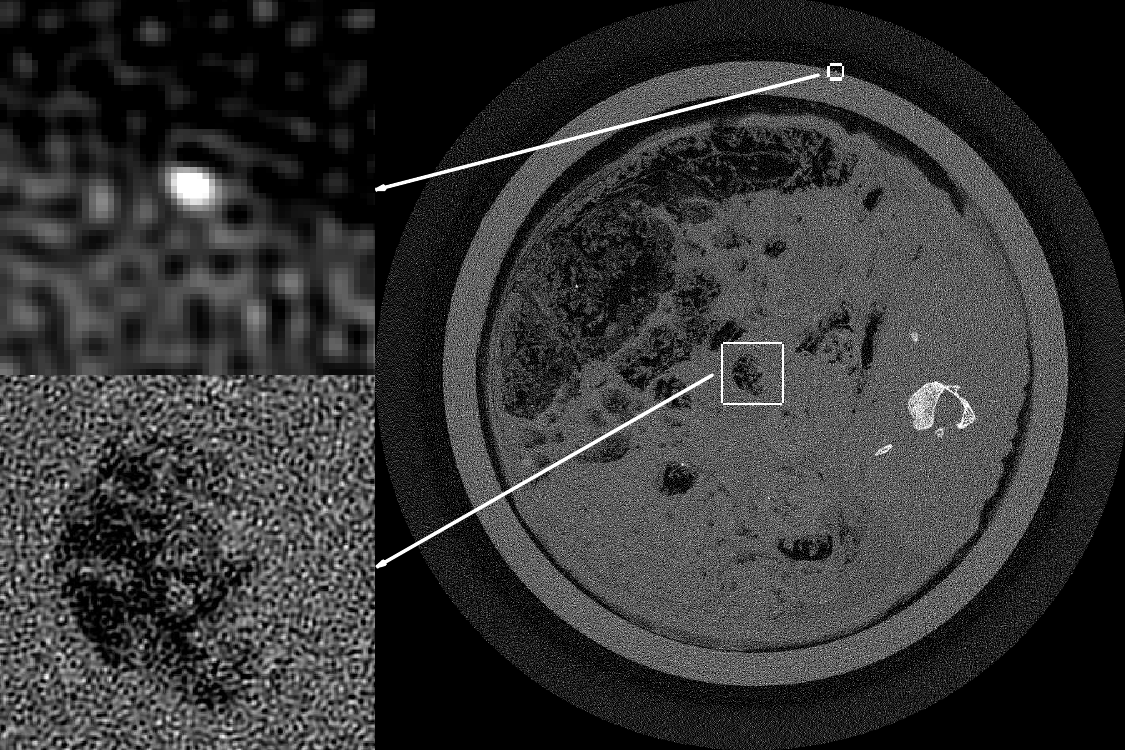}}
\end{minipage}
\caption{Unregularized FBP image reconstructed onto a 4096$\times$4096 image array.
\label{fig:FBPraw}}
\end{figure}

The main practical impact of the LASSO-type ASD-POCS algorithm comes if there is some potential advantage over
standard fan-beam FBP.  Recall that the sampling here is high-density. 
Because the sampling is so fine, we do not
expect a dramatic improvement in image quality in going from FBP to an IIR algorithm similar to what is seen with
CS-style image reconstruction with sparse views, see e.g. \cite{sidky2008image}.
Instead we expect possible improvements in image quality 
on the order of tens of percent.
A word about FBP is in order here.  The fan-beam FBP algorithm employed involves no
rounding of the ramp filter, and the corresponding unregularized image
is shown in Fig. \ref{fig:FBPraw}. The TV of this fan-beam FBP reconstruction,
denoted $t_\text{FBP}$,
is computed as a reference value for the ASD-POCS algorithm. Image reconstruction with ASD-POCS is performed for
values $t_0=t_\text{FBP}/2$, $t_\text{FBP}/4$, $t_\text{FBP}/8$, and $t_\text{FBP}/16$. As $t_0$ is decreased,
one can expect that the noise level in the image will be lower. To find a counterpart FBP image, we smoothed
the unfiltered image with a Gaussian kernel, where the kernel width is selected so that the wire amplitude matches
the corresponding ASD-POCS image. The widths of the Gaussian kernels found in this way turn out to be
$\sigma =$ 0.5, 1.2, 2.2, and 2.8, respectively, in units of pixel widths.

\begin{figure}[!t]
\begin{minipage}[b]{0.8\linewidth}
\centering
\centerline{\includegraphics[width=8cm,clip=TRUE]{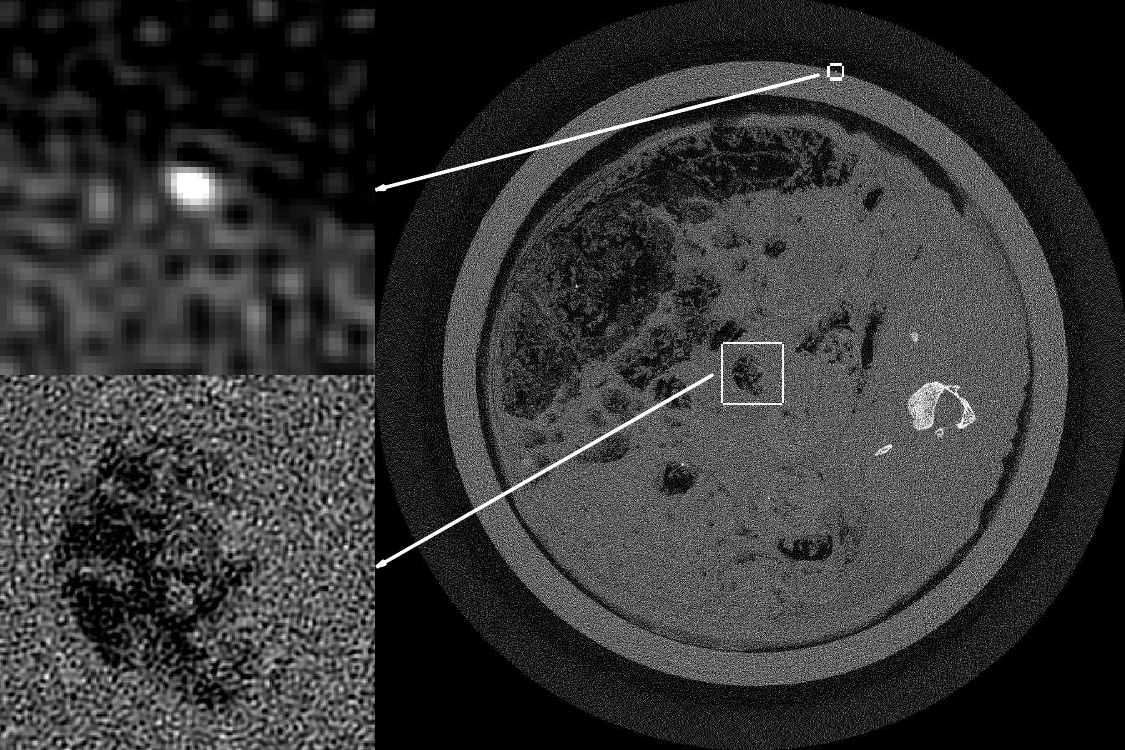}}
\vskip 0.25cm
\centerline{\includegraphics[width=8cm,clip=TRUE]{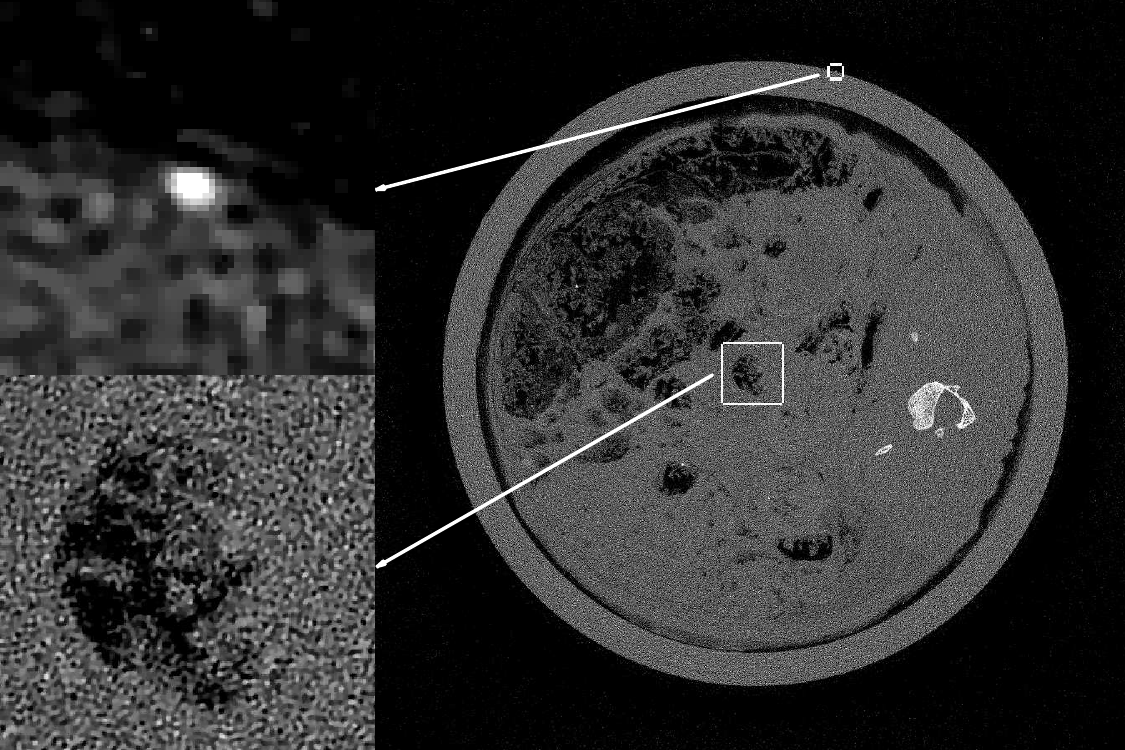}}
\end{minipage}
\caption{Top: FBP image convolved with a Gaussian of width $\sigma = 0.5$.
Bottom: ASD-POCS reconstruction for $t_0 = t_\text{FBP}/2$.
For each image the gray scale is [0, 0.06] $mm^{-1}$ accept for the top, left ROI containing the cross-section
of the wire, which is displayed in a window of [0, 0.1] $mm^{-1}$.
\label{fig:tv2}}
\end{figure}

\begin{figure}[!t]
\begin{minipage}[b]{0.8\linewidth}
\centering
\centerline{\includegraphics[width=8cm,clip=TRUE]{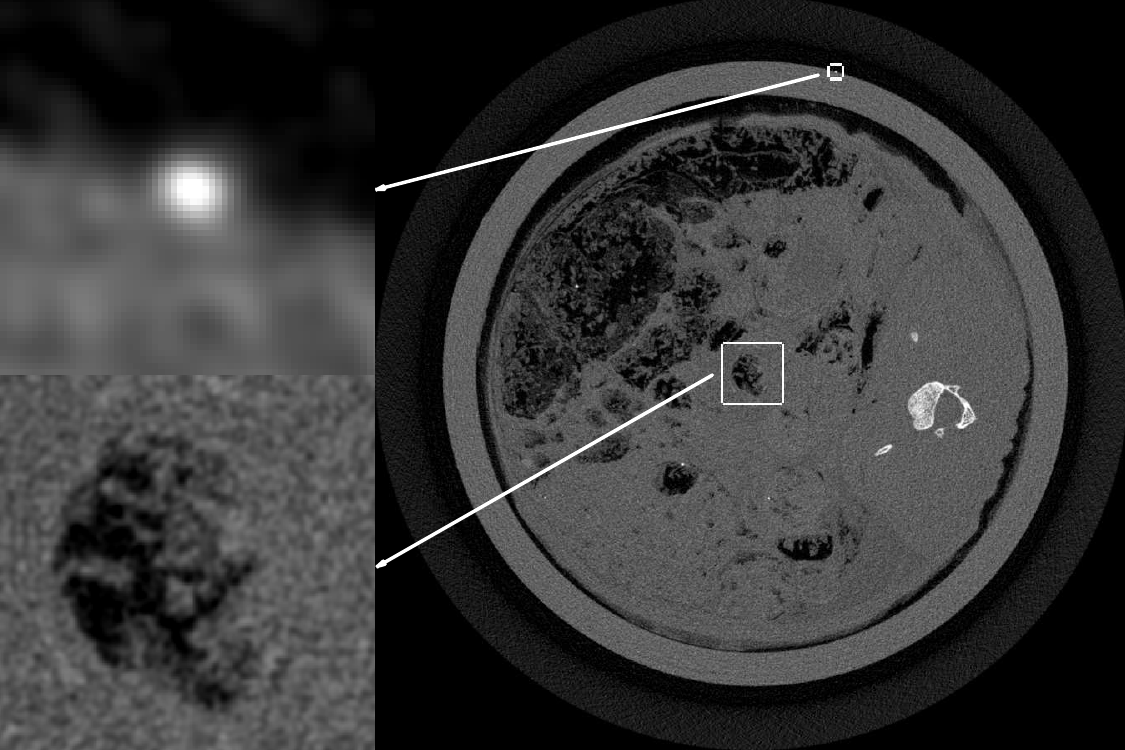}}
\vskip 0.25cm
\centerline{\includegraphics[width=8cm,clip=TRUE]{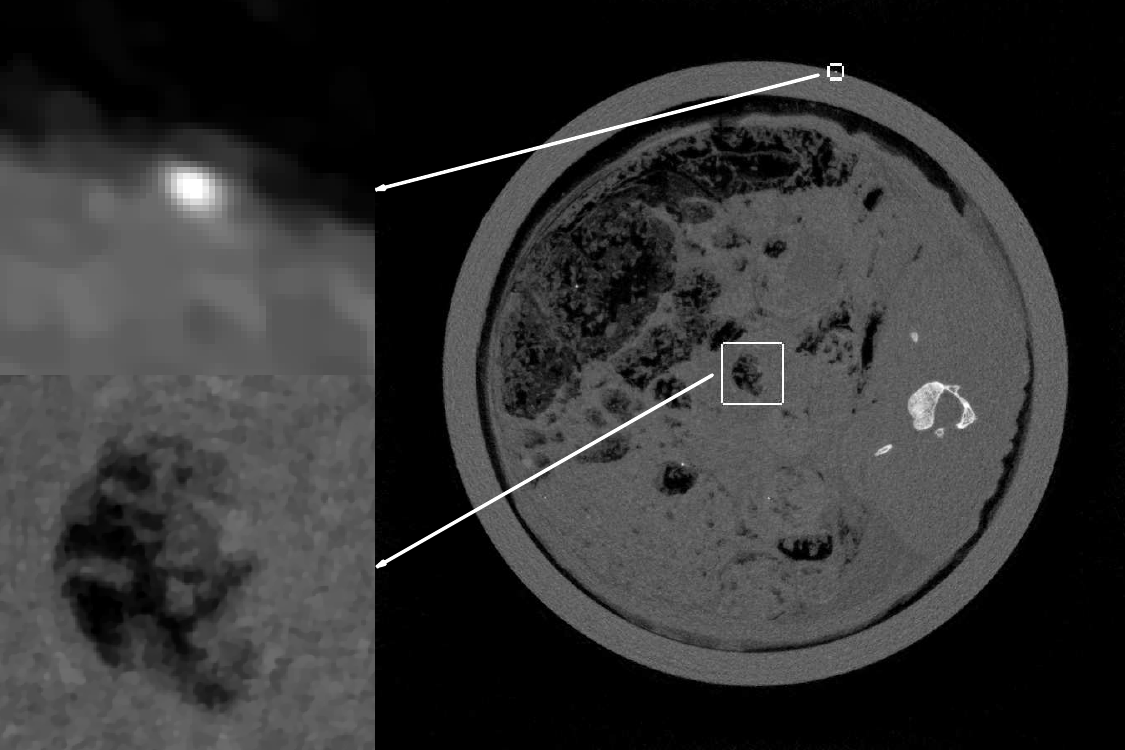}}
\end{minipage}
\caption{Top: FBP image convolved with a Gaussian of width $\sigma = 2.8$.
Bottom: ASD-POCS reconstruction for $t_0 = t_\text{FBP}/16$.
For each image the gray scale is [0, 0.06] $mm^{-1}$ accept for the top, left ROI containing the cross-section
of the wire, which is displayed in a window of [0, 0.1] $mm^{-1}$.
\label{fig:tv16}}
\end{figure}

\begin{figure}[!t]
\begin{minipage}[b]{0.8\linewidth}
\centering
\centerline{\includegraphics[width=8cm,clip=TRUE]{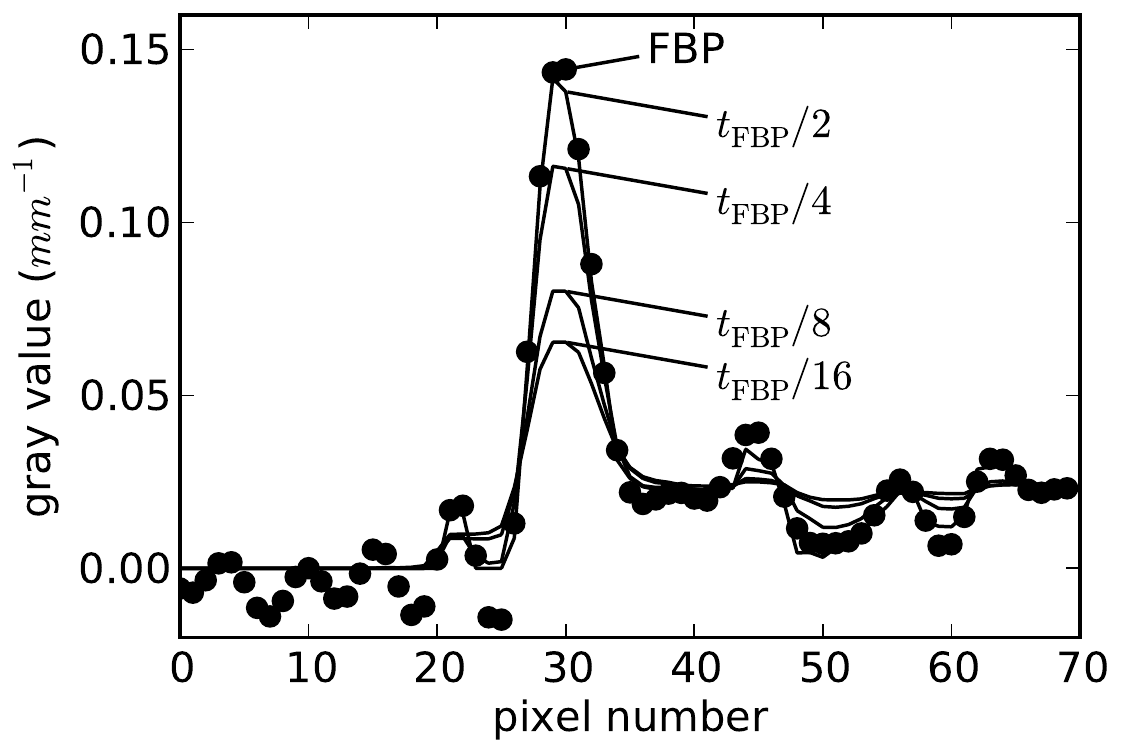}}
\end{minipage}
\caption{The solid curves represent the wire profile for ASD-POCS images for different
values of $t_0$. The dotted curve is the same profile for the unregularized FBP image.
\label{fig:TVprofs}}
\end{figure}

\begin{figure}[!t]
\begin{minipage}[b]{0.8\linewidth}
\centering
\centerline{\includegraphics[width=8cm,clip=TRUE]{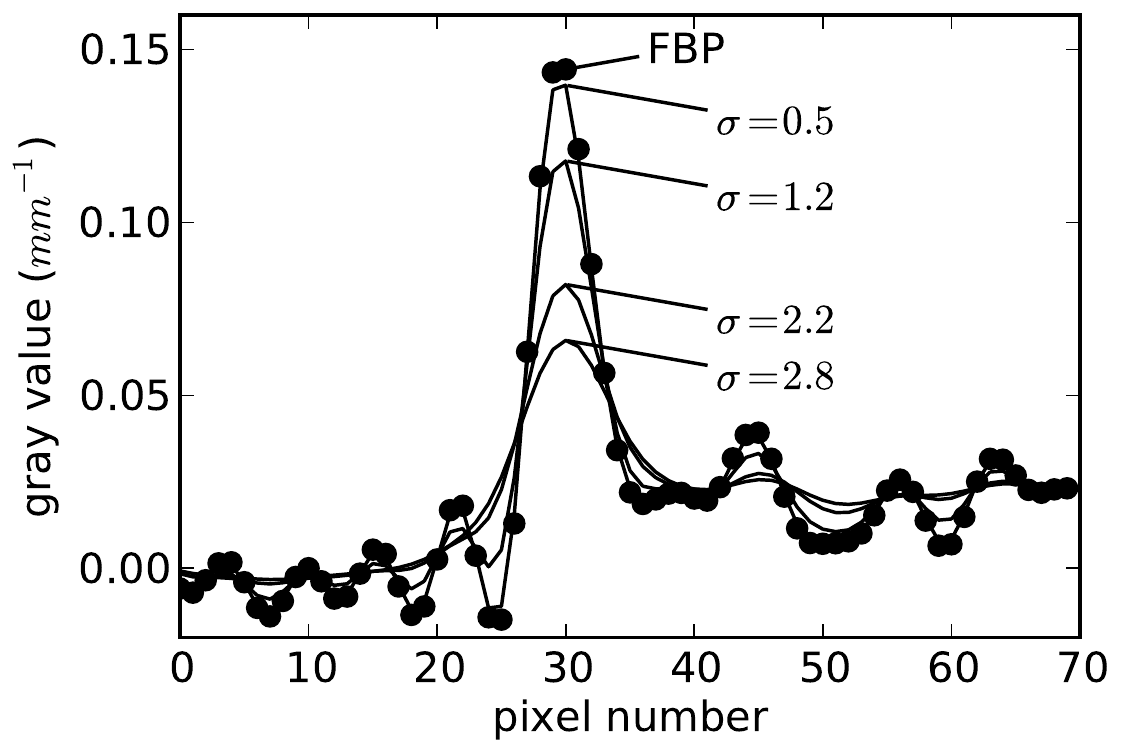}}
\end{minipage}
\caption{The solid curves represent the wire profile for FBP images smoothed by a Gaussian
of various widths.
The dotted curve is the same profile for the unregularized FBP image.
\label{fig:FBPprofs}}
\end{figure}

In Figs. \ref{fig:tv2} and \ref{fig:tv16}, we show comparisons
between the ASD-POCS images with the corresponding regularized FBP image for the least and greatest, respectively,
amount of regularization. Additionally, for more quantitative
comparison, we show a series of ASD-POCS profiles through the wire in Fig. \ref{fig:TVprofs}
and the corresponding FBP profiles in Fig. \ref{fig:FBPprofs}.

\begin{figure}[ht]
\begin{minipage}[b]{0.8\linewidth}
\centering
\centerline{\includegraphics[width=8cm,clip=TRUE]{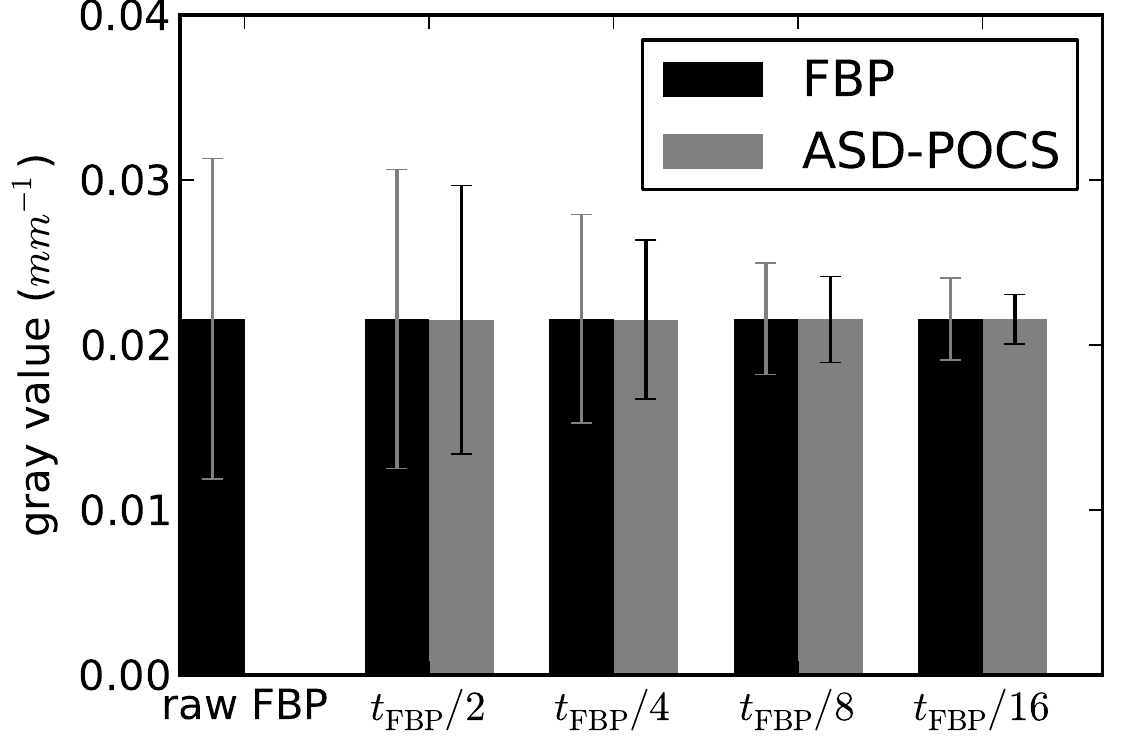}}
\end{minipage}
\caption{Mean and standard deviation over a 200$\times$200 pixel region, where the subject is
uniform. The first column shows values for the unregularized FBP image of Fig. \ref{fig:FBPraw}.
The subsequent columns are labeled by the $t_0$ value used in the ASD-POCS reconstruction.
The corresponding regularized FBP result is obtained by Gaussian smoothing where the kernel
width is selected by matching the amplitude of the wire in the ASD-POCS image, as explained in the text.
In each case the ASD-POCS yields a standard deviation lower than that of the FBP image with equivalent
contrast of the wire object.
\label{fig:barsig}}
\end{figure}

We discuss the possible advantage of IIR with the LASSO-type ASD-POCS algorithm. We point out
again and it is clear from the images that potential advantages will be small as we are trying to squeeze out
more information from a very finely sampled system.  Nevertheless, there appears to be some advantage.
Comparing Figs. \ref{fig:FBPraw} and \ref{fig:tv2}, one can see that the level of FBP regularization is
negligible as both FBP images appear very similar. The corresponding ASD-POCS image has some visible advantage,
as the noise level is lower;  this is most easily seen in the lower
left ROIs in the dark regions of the images. For all the image pairs, the noise level of the ASD-POCS image
is perceptibly lower than the corresponding FBP image. This can be seen quantitatively by computing the mean
and standard deviation of pixel values in a 200$\times$200 square just above and to the right of the bone,
where the subject's gray value is uniform.  The resulting values are displayed in a bar chart shown in
Fig. \ref{fig:barsig}.
The profile plots in Figs. \ref{fig:TVprofs} and \ref{fig:FBPprofs}
illustrate another possible advantage to the ASD-POCS algorithm. For ASD-POCS the wire profiles maintain their
width as image TV is decreased, while the Gaussian smoothed FBP profiles show spreading with increasing
regularization. This trend in the wire profile is also apparent in the 2D image of  Fig.~\ref{fig:tv16}.
In all the ASD-POCS images the smallest ROI containing
the wire cross-section still has some perceptible graininess. This graininess can be effectively removed
by further upsampling the data and reconstructing onto an 8192$\times$8192 image array. We found, however,
that the resulting gain in image quality is minimal for our purpose.

\section{Conclusion}

We have developed a CS-style image reconstruction algorithm for finely sampled projection data obtained
with a low intensity X-ray beam.  The main goals of the IIR algorithm are to provide control over the
image regularity and to image small objects of width comparable to the detector bin.  The technical
points to achieve these goals are: (1) an upsampling scheme for the projection data which takes advantage of
the asymmetry in data sampling, namely, that recognizes that the bin-direction of the data is sampled more
finely than the angular direction; and (2) conversion of the constrained, TV-minimization problem to
a LASSO formulation for the purpose of deriving an alternate ASD-POCS algorithm which efficiently solves
the corresponding optimization problem to a high degree of accuracy.  The resulting algorithm appears
to achieve the above mentioned goals.

Anecdotally, there have been complaints from radiologists that IIR images yield unrealistic looking images,
which has been blamed on the different noise patterns from IIR and FBP algorithms. We speculate that the
real issue is that IIR algorithms implemented on commercial scanners reduce the image resolution to
gain in noise reduction in a way that is difficult to control. Objects of size on the order of the detector
bin are highly distorted in standard IIR implementations. The presented ASD-POCS algorithm allows for more
control over this trade-off. We point out that the upsampling idea can be used with in conjunction with any IIR
algorithm -- a subject for future investigation.  Another direction which the current work can be extended is the
inclusion of more physics of the imaging process in the LASSO optimization problem; for example, a data error
term could be designed to more closely match the noise model of this CT system.

Addressing now the main practical issue of dose reduction while maintaining image quality, we have developed
an IIR algorithm for the extreme where IIR should have the least impact -- namely, fine sampling in the projection
angle. Fixing the overall dose, but decreasing the number of views should result in equal or better image quality
for ASD-POCS as it is originally designed for sparse-view sampling. Thus, there is a potential not only to reduce
dose, but to eliminate the need for expensive flying focal-spot technology on the X-ray source \cite{XiaFFS:09}.
This point, however,
is presently speculation as it requires a more in-depth study on data sets with similar exposure and different
numbers of projections, and there may be an additional practical issue from blurring if the X-ray source
moves at a constant rotation rate with fewer sampling intervals.

\begin{acknowledgments}

This work was supported in part by NIH R01 grants CA120540 and EB000225.
The contents of this article are
solely the responsibility of the authors and do not necessarily
represent the official views of the National Institutes of Health.
\end{acknowledgments}

\bibliographystyle{apsrev}
\bibliography{lowdoseNewer}

\end{document}